\documentstyle[preprint,aps]{revtex}
\begin{document}
\bibliographystyle{plain}
%\hskip -2\wd0\copy1
\title{ 
Magnetic properties of NaV$_2$O$_5$, a one-dimensional spin 1/2 antiferromagnet
with finite chains
}
\vskip0.5truecm 
\author {Fr\'ed\'eric Mila} 
\vskip0.5truecm
\address{
      Laboratoire de Physique Quantique, Universit\'e Paul Sabatier\\
      31062 Toulouse (France)\\
      }
\author {Patrice Millet, Jacques Bonvoisin} 
\vskip0.5truecm
\address{
	CEMES, CNRS, 29, rue J. Marvig, 31055 Toulouse Cedex (France)\\
      }
      
\maketitle

\begin{abstract}
We have performed measurements of the magnetic susceptibility 
of NaV$_2$O$_5$ between 2 and 400 K. The high temperature part is typical of 
spin 1/2 chains with a
nearest--neighbour antiferromagnetic exchange integral $J$ of 529 K. 
We develop a
model for the susceptibility of a system with finite chains to account for 
the low 
temperature part of the data, 
which cannot be fitted by a standard Curie-Weiss term. These results suggest
that the next nearest--neighbour exchange integral $J_2$ in
CaV$_4$O$_9$  should
be of the order of 500 K because, like $J$ in NaV$_2$O$_5$, it 
corresponds to corner sharing 
VO$_5$ square pyramids.
\end{abstract}
\vskip .1truein
 
\newpage 

The report by Taniguchi et al\cite{taniguchi} of a spin gap behaviour in the
quasi-two dimensional system CaV$_4$O$_9$ has triggered an intensive
theoretical activity aimed at understanding the origin of this
gap\cite{katoh,ueda,sano,albrecht,troyer,starykh,gelfand}. The emerging
picture is that there is no spin gap in the model with only 
exchange integrals $J_1$ 
between nearest neighbours\cite{albrecht,troyer,gelfand}, and that there is 
a spin gap
if a coupling constant to second neighbours $J_2$ is included as long as 
$0.2 \le J_2/J_1\le 0.7$\cite{starykh,sano,albrecht2}. To check this theory, 
one needs information on the
value of the exchange integrals. This information turns out to be difficult to
extract from the suceptibility. The best calculation of the
temperature dependence of the susceptibility of that model is a high 
temperature expansion due to Gelfand et al\cite{gelfand}. Assuming 
$J_2/J_1=1/2$, they could
reproduce the maximum of the susceptibility around 100 K with $J_1\simeq 200$ K.
The fit of the high temperature part is not satisfactory however, and the
question of the value of the integrals is still pretty much open. 

CaV$_4$O$_9$ is actually a member of a large family of vanadium oxides studied
by Galy and coworkers in the mid seventies\cite{galy}, and a natural idea is 
to look at 
other
members of the family to try to get information on the exchange integrals. The
other 2D compounds that can be synthesized with Ca, CaV$_2$O$_5$ and 
CaV$_3$O$_7$, lead to a similarly difficult problem because they involve both
$J_1$ and $J_2$. This difficulty can be overcome by studying an other mixed
valence vanadium
oxide, NaV$_2$O$_5$. 
This compound, first synthesized by Hardy et
al\cite{hardy}, is isostructural to CaV$_2$O$_5$. Note however that NaV$_2$O$_5$
crystallizes with the orthorhombic non centro-symmetric space group P2mn, while
CaV$_2$O$_5$ crystallizes with the centro--symmetric space group Pmmn.
Now, NaV$_2$O$_5$
contains Na$^+$ instead of Ca$^{2+}$, and half the vanadium have to be in the
oxydation state
V$^{5+}$ (formally one has NaV$^{5+}$V$^{4+}$O$_5$). These ions do not carry 
a spin,
while 
remaining
V$^{4+}$ carry a spin 1/2 and form a set of well separated chains of corner 
sharing VO$_5$ square pyramids (SP)
(see Fig. 1). The magnetic properties should thus be well described by the
one-dimensional spin 1/2 Heisenberg model:
\begin{equation}
H=J \sum_i \vec S_i.\vec S_{i+1}
\label{h}
\end{equation}
Note that the exchange integral $J$ between corner sharing VO$_5$ SP is
equivalent to the next-nearest neighbour exchange integral $J_2$ of
CaV$_4$O$_9$. 

In this Letter, we present measurements of the magnetic susceptibility of 
NaV$_2$O$_5$ 
from 2 to 400 K. The compound was prepared starting from a stoichiometric 
mixture of NaVO$_3$ (Merck,
min 99\%), V$_2$O$_3$ (obtained by hydrogen reduction of V$_2$O$_5$ at 800
$^{\rm o}$C) 
and V$_2$O$_5$ (Aldrich Chem. Co.,99,9\%). 
The mixture was ground intimately, sealed in an
evacuated quartz tube and then heated for 10 hours at 615 $^{\rm o}$C. 
The X-ray
diffraction pattern of the resulting dark powder indicated the formation of
the pure phase NaV$_2$O$_5$. A refinement of the structure of 
NaV$_2$O$_5$ 
was proposed by Carpy and Galy\cite{carpy1}. It is orthorhombic and consists,
as shown on the perspective
view in fig. 1a), of two dimensionnal layers of VO$_5$ SP
with the Na atoms between the layers. It is worth mentionning in this
structure the
ordering of the V$^{4+}$ and V$^{5+}$ atoms in the layers with formation of 
rows
(Fig. 1b)).
Magnetic susceptibility measurements were performed using a SQUID susceptometer.
The magnetic field intensity was 1kG. The molar susceptibilities were corrected
for diamagnetism by using Pascal's constants.

The raw data are presented in 
Fig. 2. They agree with the early
measurements between 80 and 600 K by Carpy et al\cite{carpy}.
Above 100 K, the suceptibility is consistent with that of a
spin 1/2 chain\cite{bonner,eggert}. In that temperature range, the best
available estimate of the susceptibility due to Eggert et al\cite{eggert} 
is actually
indistinguishable from the Bonner-Fisher result\cite{bonner}, so it does 
not matter which
theory we use to fit the data. There is a maximum at 350 K which implies 
an exchange integral  $J\simeq 529$ K. 

Below that temperature, there is no evidence of a phase transition or of three
dimensional ordering, but,
as usual, there is an increase of the susceptibility at low temperature
due to
some kind of defects. The standard procedure is to describe these defects by
a Curie-Weiss term $\chi^{CW}(T)=g^2\mu_B^2S(S+1)/3k_B(T-\theta)$, so that 
the spin part of the susceptibility reads:
\begin{equation}
\chi^{tot}(T)=(1-\rho)\chi^\infty (T) + \rho \chi^{CW} (T) + \chi^{VV}
\label{chi}
\end{equation}
$\chi^{VV}$ is the temperature independent Van Vleck paramagnetic 
susceptibility,
$\rho$ is the concentration of impurities, and $\chi^\infty (T)$ is the
susceptibility of the infinite chain. At low temperature, the difference between
the Bonner-Fisher estimate and the recent results of
Eggert et al\cite{eggert} is not negligible\cite{eggert2}, 
and we have used the results of Eggert et al for $\chi^\infty (T)$.  
It turns out that the low temperature part of the susceptibility cannot be 
fitted
satisfactorily along these lines. The best fit one can get using Eq. 
(\ref{chi}) is depicted as a dashed line in Fig. (2). It considerably 
overestimates the actual susceptibility around the minimum at 70 K. 

The main problem is that the amount of impurities one needs to interpret the low
temperature susceptibility gives a much too large contribution at higher
temperatures. In other words,
the susceptibility behaves as if the impurities were slowly disappearing when
the temperature increases. While this clearly cannot be reconciled 
with extrinsic impurities, such a behaviour actually makes sense if the 
impurity contribution comes from finite chains with an odd number of sites. 
The idea is the following: Roughly speaking, a
finite-length chain with N spins behaves like an infinite one at
temperatures larger than the finite-size gap, and like a finite one below that
temperature. Now, the finite size gap is of order $J/N$. So if we have a
distribution of finite chains with different lengths, they will progressively
dissappear from the impurity term to contribute to $\chi^\infty (T)$ as the
temperature is increased.

To be more quantitative, we need to know the distribution of length of the
finite chains. If we make the reasonable assumption that the finite chains
are due to a random distribution of point defects, then elementary statistical
mechanics shows that the distribution of length is of the form
\begin{equation}
P(N)={\exp(-\lambda N) \over Z}
\label{P}
\end{equation}
where the partition function $Z=(1-\exp(-\lambda))^{-1}$ while the
Lagrange parameter
$\lambda$ is fixed by the average number of sites $N_0$ of the chains 
according to
$\lambda = \log (1+1/N_0) \simeq 1/N_0$ if $N_0\gg 1$. The average number of
sites of 
the
chains is itself related to the concentration $\rho$ of defects by $N_0=1/\rho$.
Then
the total susceptibility per spin is simply given by
$\chi(T)=(1/N_0)\sum_N P(N) \chi_N(T)$ where $\chi_N(T)$ is the susceptibility 
of a
chain of length $N$.

Numerical calculations of $\chi_N(T)$ are indeed possible\cite{bonner},
but even with today's numerical
facilities, accurate results would be limited to relatively short chains.
However, a good estimate can be obtained in the following way. 
At low enough temperature, a finite chain with an odd number of sites
behaves like a spin 1/2 impurity according to Curie's law
$\chi^{Curie}(T)=g^2\mu_B^2S(S+1)/3k_BT$,
while the susceptibility of a chain with an even number of sites vanishes
exponentially. Averaging the susceptibility according to
$\chi_N^{av}=(\chi_N+\chi_{N+1})/2$, and neglecting the exponentially small
contribution of the even chain, the susceptibility of finite chains is given by
$\chi_N^{av}(T)=\chi^{Curie}(T)/2$ at low temperature and
$\chi_N^{av}(T)=N\chi^\infty(T)$ at high temperature. So, if we define the 
cross--over 
temperature $T_c(N)$ as the temperature where these two
expressions are equal, we can approximate $\chi_N(T)$ by 
$\chi^{Curie}(T)/2$ if $T<T_c(N)$ and $N\chi^\infty(T)$ if $T>T_c(N)$. 
Equivalently, one can define a length scale $L(T)$ by
$k_BT_c(L(T)/a)=J$,
where $a$ is the lattice parameter.
Then, for a given temperature $T$, $\chi_N(T)$ will be given by 
$N\chi^\infty(T)$
if $N>L(T)/a$ and by $\chi^{Curie}(T)/2$
if $N<L(T)/a$. 
Solving $k_BT_c(L(T)/a)=J$ for $L(T)$ yields
$L(T)/a=\chi^{Curie}(T)/2\chi^\infty(T)$. This expression 
can be cast in a more transparent form if
one writes $\chi^\infty(T)={(g\mu_B)^2 \over J} \bar \chi(T)$, where $\bar
\chi(T)$ is the normalized susceptibility that equals $1/\pi^2$ at $T=0$. In
terms of this function, one has $L(T)/a=(1/8\bar \chi(T))J/k_BT$. Note that 
$8\bar \chi(T)$ is of order 1 for $T<J$. The cross-over
temperature $T_c(N)\simeq J/N$ is thus of the order of the finite size gap, 
as it should. 

So, the total susceptibility per site is given by
\begin{equation}
\chi(T)=(1/N_0)\left( {\chi^{Curie}(T) \over 2} \sum_{N<L(T)/a} P(N) 
+ \chi^\infty(T)\sum_{N>L(T)/a} N\ P(N) \right)
\end{equation}
This equation is actually valid for any distribution of chain length $P(N)$.
Concentrating on the distribution of Eq. (\ref{P}),
the sums are readily performed: $\sum_{N<L(T)/a}
P(N)=1-\exp(-L(T)/aN_0)$ and $\sum_{N>L(T)/a} N\ P(N)=
N_0-(N_0+L(T)/a)\exp(-L(T)/aN_0$. Including a VanVleck contribution,
our final result for the susceptibility reads
\begin{equation}
\chi^{tot}(T)={1-\exp(-L(T)/aN_0) \over 2N_0} \chi^{CW} (T) + 
\left( 1+{L(T) \over aN_0} \right) \exp(-L(T)/aN_0) \chi^\infty(T) + \chi^{VV}
\label{chi2}
\end{equation}
with $L(T)/a=(1/8\bar \chi(T))J/k_BT$\cite{note}. We have replaced
 $\chi^{Curie}(T)$ by
$\chi^{CW}(T)$ to account for possible residual interactions between the finite
chains. This can be seen as an extension of Eq. (\ref{chi}).
The concentration $\rho=1/N_0$ is still an adjustable parameter. However, the
coefficients in front of $\chi^{CW} (T)$ and $\chi^\infty(T)$ are no longer
simply $\rho$ and $1-\rho$, but functions of temperature.
Using Eq. (\ref{chi2}), we have been able to obtain a much better fit of
the raw experimental data. This fit is shown as a solid line on Fig. (2). 
The parameters are $J=529$ K, $g=2.043$, $\chi^{VV}=141.
10^{-6}ccm/mole$, $N_0=35$ and $\theta =-1.26$ K. 
This value of $N_0$ 
corresponds to a concentration
of defects $\rho=2.9$ \% which seems to be a reasonable number. 
The fit was realized with the help of a simplex non-linear least-squares fitting
procedure, and the relative deviation defined as
$\sum_i(\chi^{obs}_i-\chi^{calc})^2/\sum(\chi^{obs}_i)^2$ was equal to
$3.10^{-5}$. 

Let us come back to CaV$_4$O$_9$ for a moment.
Our data for the susceptibility of NaV$_2$O$_5$ are consistent with the 
model of Eq.
(\ref{h}) with an exchange constant $J=529$ K. This value should be contrasted
with the values proposed for $J_2$ in the case of CaV$_4$O$_9$, which 
range from 50
to 100 K\cite{ueda,starykh,gelfand}. While one cannot exclude some dependence 
of the exchange integrals on
the overall chemical environment, they should essentially depend on the local
geometry, which is the same in NaV$_2$O$_5$ and CaV$_4$O$_9$ for corner sharing
SP. Besides, the present determination of $J$ is quite unambiguous 
because this is the only exchange integral involved. So we are led to the
conclusion that previous estimates of $J_2$ in CaV$_4$O$_9$ cannot be correct.
If we put all the information we have at the moment, a consistent picture of
exchange integrals can still be obtained. The presence of a gap of 107 K in 
CaV$_4$O$_9$, i.e. much smaller than $J_2$, can still be explained by the
$J_1-J_2$ model on the depleted lattice if $J_2/J_1$ is not too far from either
of the
critical values 0.2 and 0.7 where the gap disappears. Now, according to a recent
work of Kontani et al\cite{kontani}, the observation by neutron scattering of 
stripe order 
in the compound CaV$_3$O$_7$\cite{harashina} implies that $J_2/J_1$ cannot be 
too small. At a
quantitative level, the
bound given by the modified spin-wave theory $J_2/J_1>0.6932$
cannot be taken too seriously, but a ratio $J_2/J_1$ close to 0.2 can be
excluded because CaV$_3$O$_7$ should then exhibit N\'eel order. So we think
that coupling constants $J_1\simeq 700$ K and $J_2\simeq 500$ K are the best
candidates so far to describe this type of vanadium oxides. A very useful check
will be to see if such values are compatible with the temperature 
dependences of the susceptibility reported for CaV$_3$O$_7$\cite{liu} and 
CaV$_4$O$_9$\cite{taniguchi}.

In conclusion, we have developped a theory to describe the low temperature
susceptibility of a one-dimensional spin 1/2 antiferromagnetic system with
finite chains. This theory leads to a very nice fit of the data we have obtained
for NaV$_2$O$_5$. More generally, this theory should provide a much more
accurate way of substracting the contribution of finite chains from the
experimental data than just adding a Curie-Weiss term. 
This might help identify in other compounds the remarkable 
temperature dependence
predicted by Eggert et al\cite{eggert} for the infinite spin 1/2
antiferromagnetic chain and already observed in Sr$_2$CuO$_3$ according to
Eggert's reinterpretation of the experimental data\cite{eggert2}. 
It would be quite interesting to see how accurate the present theory is by
performing 
Monte Carlo calculations of the temperature dependence of the susceptibility of 
finite chains. 
It would also be very interesting to test it
on
systems where the concentration of non
magnetic defects that break chains can be controlled. Work along these lines on
various
vanadium based oxides is in progress.

We acknowledge useful discussions with M. Albrecht, J. P. Daudey,  
M. Luchini, D.
Poilblanc,
J.-P. Renard and J. M. Savariault. We are especially grateful to J. Galy for
sharing with us his expertise on vanadium oxides.
We are also grateful to S. Eggert for
sending us his numerical estimates of the susceptibility of 
Ref. \cite{eggert}.

\begin{figure}
\caption{Structure of NaV$_2$O$_5$. a) Perspective view in the [010] direction.
The square pyramids occupied by V$^{4+}$ are indicated by an arrow. b)
Schematic representation of the (V$_2$O$_5$)n rows along the Oy axis.
}
\end{figure}

\begin{figure}
\caption{Thermal variation of the molar magnetic susceptibility of 
NaV$_2$O$_5$. Diamonds:
Experimental data; Broken line: Fit using Eq. (2); Solid line: Fit using the
present theory (Eq. 5). Insert: Enlargement of the low temperature region.
}

\end{figure}

\begin{references}

\bibitem{taniguchi} S. Taniguchi, T. Nishikawa,Y. Yasui, U.Kobayashi, M. Sato,
T. Noshioka, M. Kontani and K. Sano, J. Phys. Sos. Jpn. {\bf 64}, 2758 (1995).

\bibitem{katoh} N. Katoh and M. Imada, J. Phys, Soc. Jpn. {\bf 64}, 4105 (1995).

\bibitem{ueda} K. Ueda, H. Kontani, M. Sigrist and P. A. Lee, Phys. Rev. Lett. 
{\bf 76}, 1932 (1996).

\bibitem{sano} K. Sano and K. Takano, J. Phys, Soc. Jpn. {\bf 65}, 46 (1996).

\bibitem{albrecht} M. Albrecht and F. Mila, Phys. Rev. B {\bf 53}, R2945 (1996).

\bibitem{troyer} M. Troyer, H. Kontani and K. Ueda, unpublished.

\bibitem{starykh} O. A. Starykh, M. E. Zhitomirsky, D. I. Khomskii, R. R. P.
Singh and K. Ueda, unpublished.

\bibitem{gelfand} M. Gelfand, Z. Weihong, R. R. P. Singh, J. Oitmaa and C. J.
Hamer, unpublished.

\bibitem{albrecht2} M. Albrecht, F. Mila and D. Poilblanc, unpublished.

\bibitem{galy} See J. C. Bouloux and J. Galy, J. Solid State Chem. {\bf 16}, 385
(1976), and references therein.

\bibitem{hardy} 
A. Hardy, J. Galy, A. Casalot and M. Pouchard, Bull. Soc. Chim. Fr.
{\bf 4}, 1056 (1965);
J. Galy, A. Casalot, M. Pouchard, P. Hagenmuller, C. R. Acad. Sc.
{\bf 262 C}, 1055 (1966);
M. Pouchard, A. Casalot, J. Galy et P. Hagenmuller,Bull. Soc.
Chim. Fr. {\bf 11}, 4343 (1967).

\bibitem{carpy1} A. Carpy and J. Galy, Acta Cryst. {\bf B31}, 1481 (1975).

\bibitem{carpy} A. Carpy, A. Casalot, M. Pouchard, J. Galy and P. Hagenmuller,
J. Solid State Chem. {\bf 5}, 229 (1972).

\bibitem{bonner} J. C. Bonner and M. E. Fisher, Phys. Rev. {\bf 135}, A640
(1964).

\bibitem{eggert} S. Eggert, I. Affleck and M. Takahashi, Phys. Rev. Lett. {\bf
73}, 332 (1994).

\bibitem{eggert2} S. Eggert, unpublished.

\bibitem{note}
For practical purposes, the function $\bar \chi(T)$ can be approximated
by the following analytical expression ($x=T/J$): a) $x<0.022$:        
$\bar \chi(T)=1/\pi^2+1/(2\pi^2\log(7.3/x))$; b) $0.022<x<0.3$: 
$\bar \chi(T)=0.1084232
+0.0801571x
-0.318277x^2
+1.56832x^3
-2.00499x^4$; c) $x>0.3$: $\bar \chi(T)=0.10062+ 0.052056 x+  0.374768 x^2 -
0.997104  x^3+ 0.986688 x^4  -0.4912832  x^5+0.12330496   x^6 -0.0124234 
x^7$. The expression for $x>0.3$ is nothing but the fit of the Bonner-Fisher 
result proposed by Torrance et al (Phys. Rev.
B {\bf 15}, 4738 (1977)) with the appropriate rescaling
of J by a factor 2.

\bibitem{kontani} H. Kontani, M. E. Zhitomirsky and K. Ueda, unpublished.

\bibitem{harashina} H. Harashina, K. Kodama, S. Shamoto, S. Taniguchi, T.
Nishikawa, M. Sato, K. Kakurai and M. Nishi, unpublished.

\bibitem{liu} G. Liu and J. E. Greedan, J. Solid State Chem {\bf 103}, 139
(1993).

\end{references}
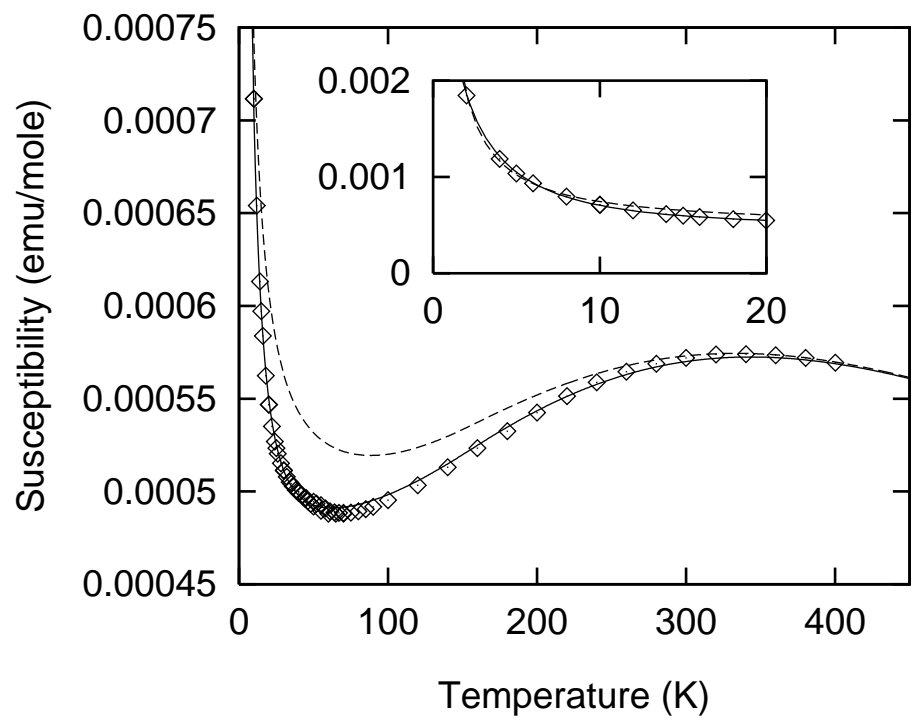
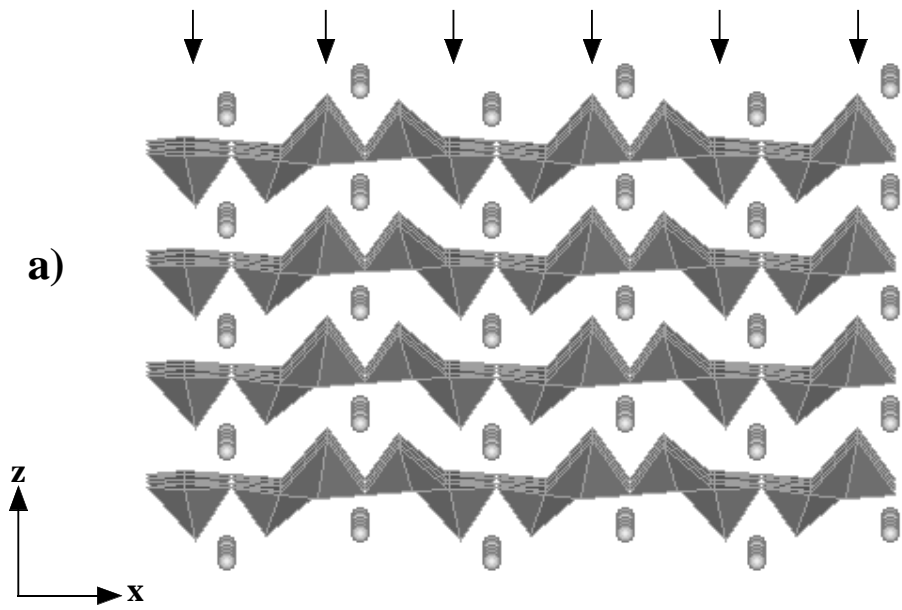
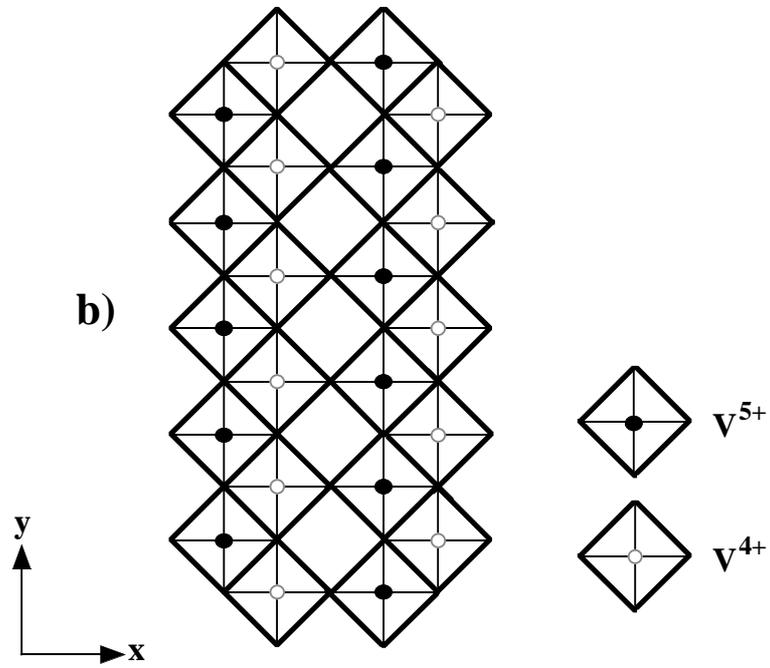
\end{document}